\documentclass[conference]{IEEEtran}
\IEEEoverridecommandlockouts

\usepackage{cite}
\usepackage{amsmath,amssymb,amsfonts}
\usepackage{graphicx}
\usepackage{textcomp}
\usepackage{xcolor}
\usepackage{url}
\usepackage{booktabs}
\usepackage{array}
\usepackage{balance}
\newcommand{\Fig}[1]{Fig.~\ref{#1}}
\newcommand{\Eq}[1]{(\ref{#1})}
\newcommand{\E}{\mathrm{E}}

\begin{document}

\title{Spaceborne SAR Change Detection and Coherence Analysis for Maritime Port Monitoring}

\author{\IEEEauthorblockN{Necati Kagan Erkek \IEEEauthorrefmark{2} and Kudret Esmer\IEEEauthorrefmark{2}}
\IEEEauthorblockA{\IEEEauthorrefmark{2} Telecommunications Engineering, Department of Electronic, Information and Bioengineering \\ Politecnico di Milano, Piazza Leonardo da Vinci 32, 20133, Milan, Italy} \texttt{necatikagan.erkek@mail.polimi.it, kudret.esmer@mail.polimi.it}}
\maketitle

\begin{abstract}
Spaceborne synthetic aperture radar (SAR) provides coherent microwave imagery suitable for maritime infrastructure monitoring under illumination-independent and weather-independent acquisition conditions. An academic conference-style analysis is presented for SAR amplitude and geocoded multitemporal data over Tianjin Port, China. The processing chain includes amplitude visualization, radiometric scaling, view-direction interpretation, range and azimuth resolution assessment, speckle reduction, amplitude-based change mapping, GeoTIFF export for geographic inspection, and interferometric coherence estimation. Histogram-guided display limits improve the interpretability of the complex SAR magnitude images, while zoomed inspection of shadows and bright layover responses supports qualitative interpretation of illumination geometry. A two-dimensional Fourier analysis is used to characterize dominant spectral content and to estimate an approximate range resolution of 0.42 m and an azimuth angular separation of 0.19 degrees under the available image-coordinate calibration. Multitemporal master and slave images are subsequently compared through filtered amplitude differences and coherence maps computed with multiple spatial averaging windows. The results highlight the relevance of SAR amplitude and coherence products for detecting structural and surface-condition variations in dense port environments characterized by vessels, storage tanks, quay structures, industrial yards, and water-land transitions.
\end{abstract}


\section{Introduction}

Spaceborne synthetic aperture radar (SAR) is an active coherent imaging technique that enables high-resolution observation of the Earth's surface without reliance on solar illumination and with reduced sensitivity to cloud cover and weather conditions \cite{moreira2013,curlander1991,cumming2005}. The side-looking acquisition geometry, synthetic aperture formation, and coherent phase history allow SAR systems to resolve maritime, urban, agricultural, and natural targets at spatial scales that are highly relevant for operational Earth observation \cite{franceschetti1999,oliver1998,lee2009}. Maritime ports constitute demanding SAR scenes because metallic vessels, quay walls, cranes, tanks, roads, railways, and industrial roofs generate strong scattering, while adjacent water areas often exhibit low backscatter, wave-dependent texture, and specular behavior \cite{oliver1998,lee2009}. Multitemporal SAR analysis is widely used for detecting land-cover, infrastructure, and surface-condition changes. Amplitude-based methods commonly rely on differencing, ratioing, log-ratioing, statistical thresholding, or model-based similarity measures, with speckle reduction remaining a central challenge \cite{rignot1993,bruzzone2000,inglada2007,moser2006}. Classical adaptive filters and local-statistics filters reduce multiplicative speckle while attempting to preserve edges and isolated bright scatterers \cite{lee1980,touzi2002}. In parallel, interferometric SAR (InSAR) uses the complex relationship between two SAR acquisitions to estimate phase and coherence, providing information on scene stability, decorrelation, topography, and deformation \cite{bamler1998,ferretti2007,zebker1992,touzi1999}.

\begin{figure}[!t]
    \centering
    \includegraphics[width=0.98\linewidth]{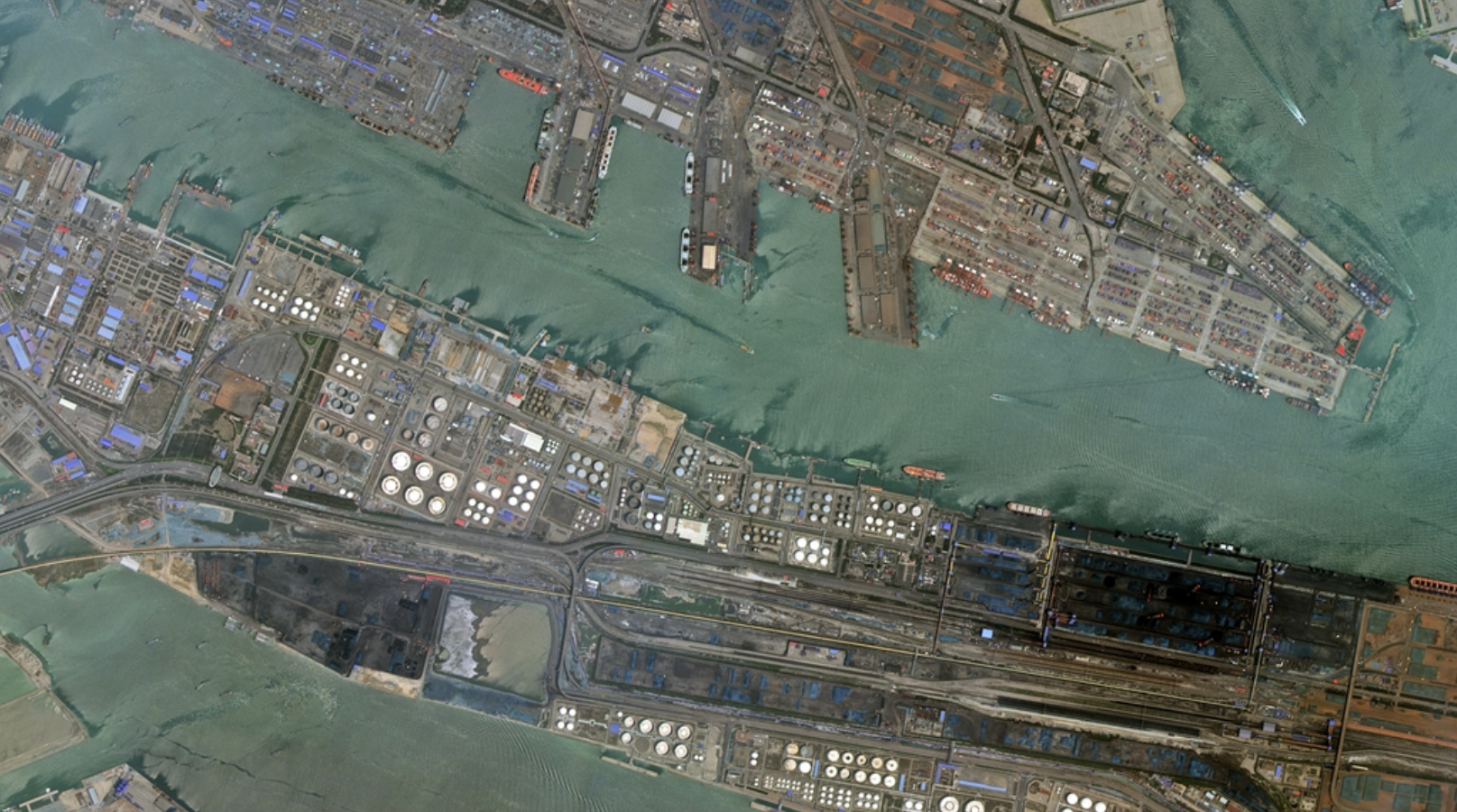}
    \caption{Optical overview of the analyzed maritime port area used as geographic context for interpreting SAR amplitude, change-detection, and coherence products.}
    \label{fig:tianjin-optical}
\end{figure}

The objective is to evaluate a compact SAR processing workflow for Tianjin Port, combining visual interpretation, resolution assessment, amplitude change detection, and coherence estimation. The analysis uses MATLAB-based processing for complex SAR arrays and geocoded products, followed by geographic inspection through exported GeoTIFF layers. An optical reference image of Tianjin Port is included as geographic context in \Fig{fig:tianjin-optical}; the optical scene illustrates the dense mixture of water, docks, vessels, road infrastructure, industrial zones, and storage-tank fields that influence the SAR responses analyzed in the following sections.

\section{Study Area, Data, and Processing Workflow}

\subsection{Port Scene Characteristics}

The study area is a highly engineered coastal port environment containing navigation basins, quays, breakwaters, road and rail corridors, container and bulk-cargo zones, industrial roofs, storage tanks, and low-backscatter water bodies. Such a scene produces strong SAR contrast. Smooth water surfaces generally return weak energy to a side-looking radar, whereas metallic structures, vertical walls, tank rims, ship hulls, and orthogonal infrastructure can generate intense single-bounce, double-bounce, or layover responses. These scattering mechanisms make the site suitable for evaluating both amplitude-based change detection and complex coherence.

The optical reference in \Fig{fig:tianjin-optical} provides spatial context for the geocoded SAR products. It is not used as a radiometric input to the SAR calculations; instead, it supports interpretation of port morphology, dock geometry, storage-tank fields, and water-land transitions. This separation is important because the optical image measures reflected solar radiation, whereas SAR measures microwave backscatter determined by surface roughness, geometry, dielectric properties, and radar-viewing direction.

\subsection{Input Data, Notation, and Software Environment}

The input data consist of a complex range-azimuth SAR image $Z\in\mathbb{C}^{N_r\times N_a}$ and a geocoded multitemporal pair denoted by master $M$ and slave $S$. The geocoded pair is accompanied by a spatial reference object $R$, allowing the processed raster layers to be exported as GeoTIFF files and examined in a geographic information system. Unless stated otherwise, image visualization is performed on magnitude images rather than on complex samples. For a complex pixel $Z_{ij}=I_{ij}+jQ_{ij}$, the amplitude is
\begin{equation}
A_{ij}=|Z_{ij}|=\sqrt{I_{ij}^{2}+Q_{ij}^{2}} .
\label{eq:amplitude}
\end{equation}
This operation removes phase information but preserves the backscatter-strength information used for display, filtering, and amplitude-difference mapping.

A reproducible processing chain is summarized in Table~\ref{tab:workflow}. The same display interval is retained for comparable amplitude products so that visual differences are not caused by inconsistent color scaling. MATLAB is used for array processing, filtering, spectral analysis, and GeoTIFF export; QGIS is used only for geographic visualization of exported rasters.

\begin{table}[!h]
\centering
\caption{Summary of the SAR processing workflow.}
\label{tab:workflow}
\begin{tabular}{p{0.28\linewidth}p{0.58\linewidth}}
\toprule
\textbf{Stage} & \textbf{Purpose} \\
\midrule
Magnitude formation & Convert complex SAR samples to amplitude images through \Eq{eq:amplitude}. \\
Radiometric display & Select robust visualization limits from amplitude histograms. \\
View-direction analysis & Interpret shadows, layover, and bright scatterers in zoomed regions. \\
Resolution assessment & Relate range and azimuth separability to bandwidth, aperture synthesis, and spectral support. \\
Speckle reduction & Apply local median filtering before qualitative change-map interpretation. \\
Change detection & Produce filtered amplitude-difference layers from the master and slave acquisitions. \\
Coherence estimation & Compute local complex correlation using several spatial averaging windows. \\
GeoTIFF export & Transfer processed rasters to a geographic information system through reference object $R$. \\
\bottomrule
\end{tabular}
\end{table}

\section{Methodology}

\subsection{Amplitude Visualization and Radiometric Scaling}

\begin{figure}[!h]
    \centering
    \includegraphics[width=0.98\linewidth]{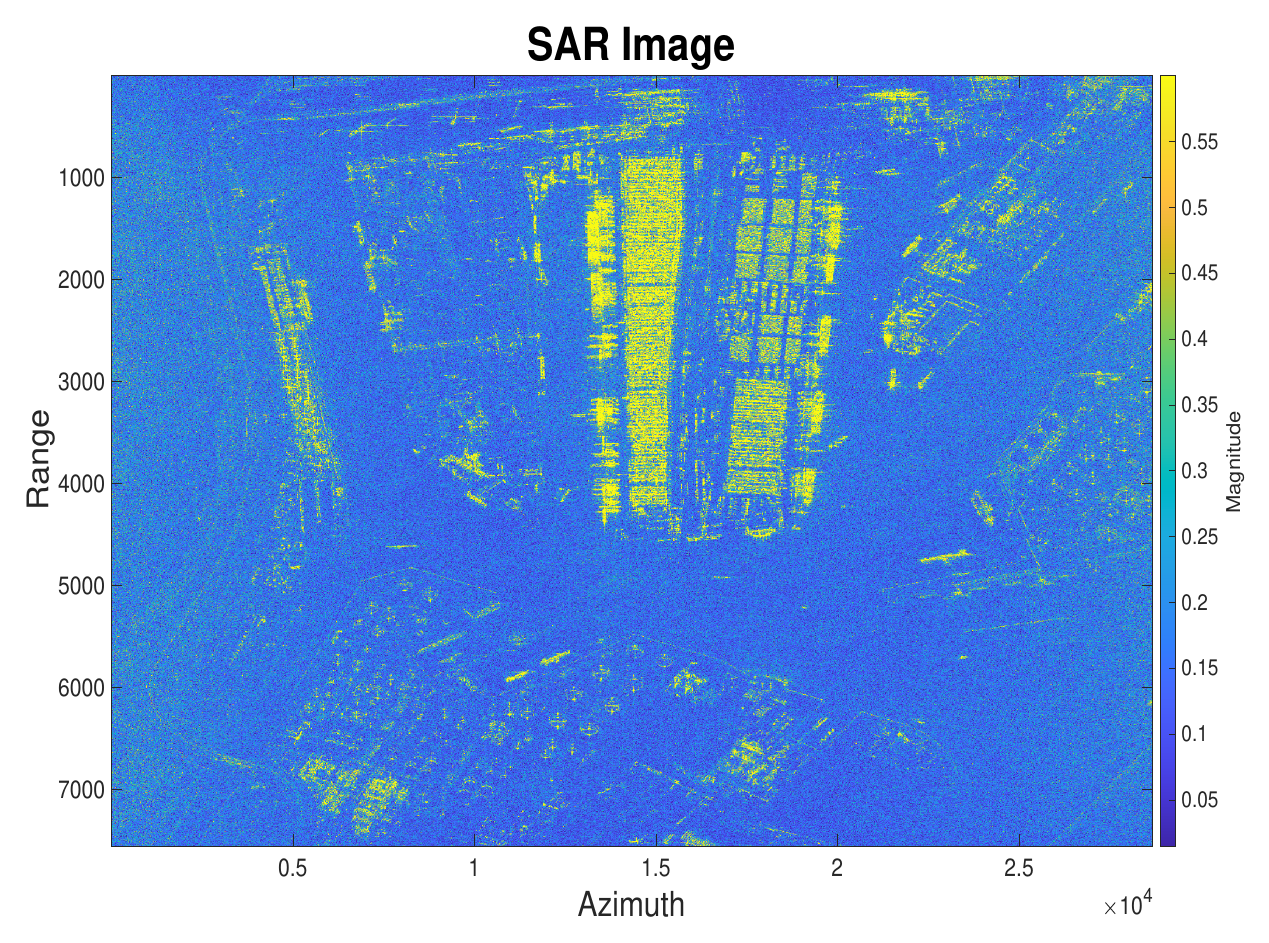}
    \caption{SAR magnitude image displayed with histogram-guided color scaling.}
    \label{fig:sar-image}
\end{figure}

\begin{figure}[!h]
    \centering
    \includegraphics[width=0.94\linewidth]{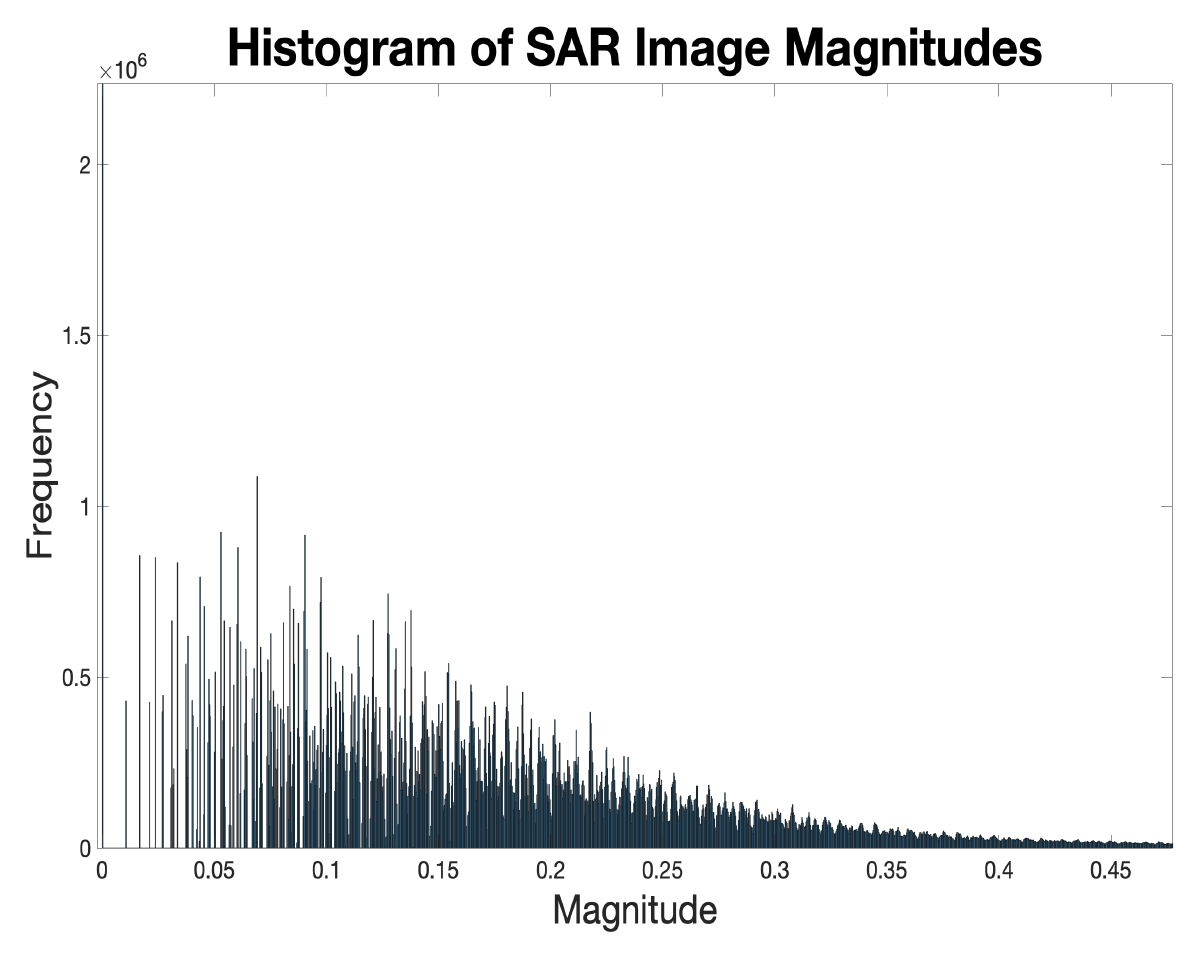}
    \caption{Histogram of SAR magnitude values used to select the display interval.}
    \label{fig:histogram}
\end{figure}

\begin{figure}[!h]
    \centering
    \includegraphics[width=0.98\linewidth]{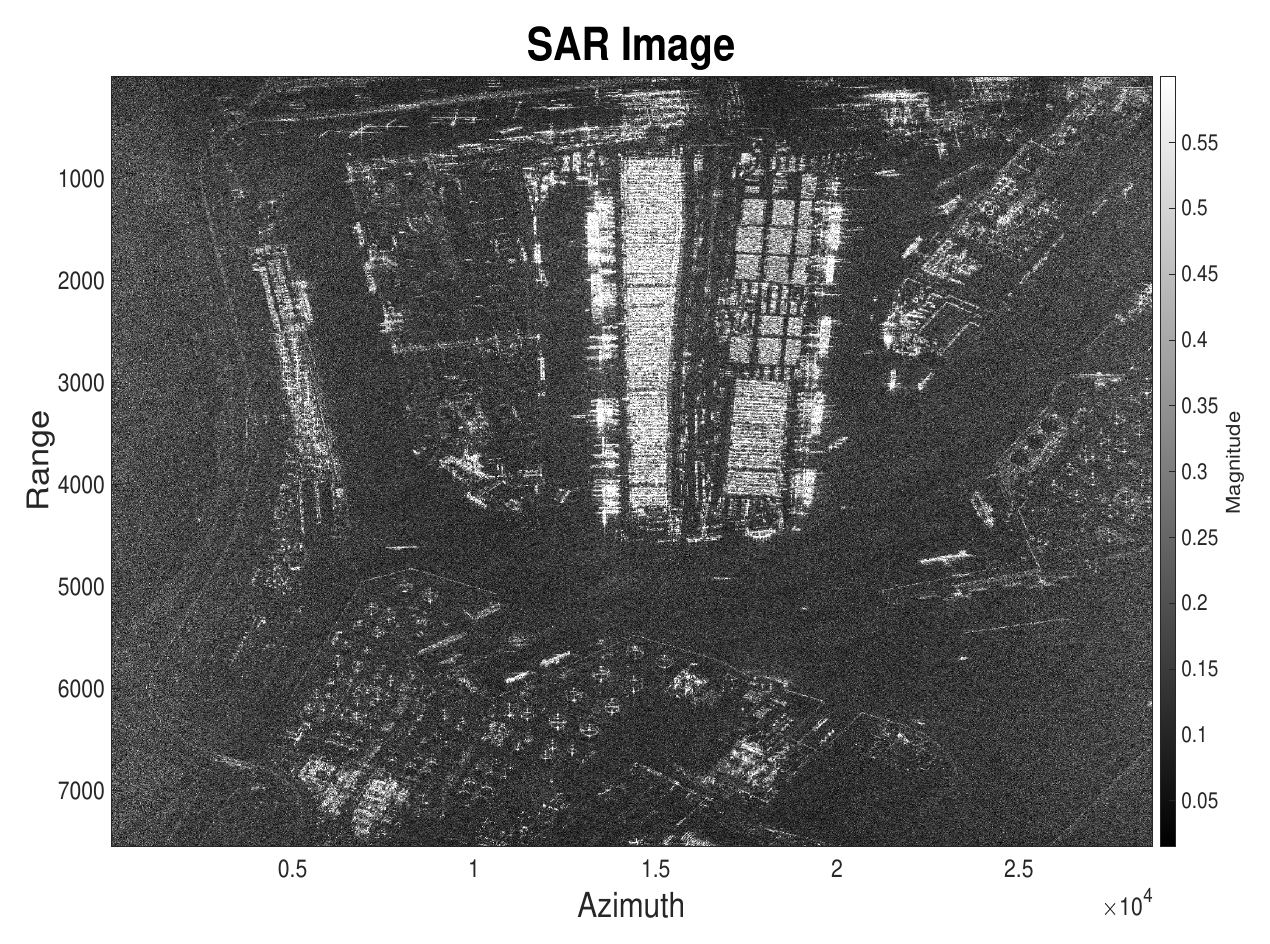}
    \caption{Grayscale SAR magnitude image after radiometric display adjustment.}
    \label{fig:gray-sar}
\end{figure}

The raw magnitude image has a heavy-tailed distribution typical of high-resolution SAR scenes. A limited number of strong scatterers can dominate the dynamic range and make moderate-amplitude port structures difficult to observe. For display purposes, an amplitude-clipped image is formed as
\begin{equation}
A_c(i,j)=\min\left\{u,\max\left[l,A(i,j)\right]\right\},
\label{eq:clipping}
\end{equation}
where $l$ and $u$ are lower and upper display limits. These limits do not change the stored SAR measurements; they only control the plotted color scale. The amplitude histogram supports the selection of $l=0$ and $u=0.6$ for the analyzed data. The color display, histogram, and grayscale rendering are shown in \Fig{fig:sar-image}, \Fig{fig:histogram}, and \Fig{fig:gray-sar}, respectively. The consistent interval improves comparability between subsequent figures and prevents isolated bright reflectors from suppressing weaker but meaningful spatial patterns.

\subsection{View-Direction Interpretation}

\begin{figure}[!h]
    \centering
    \includegraphics[width=0.98\linewidth]{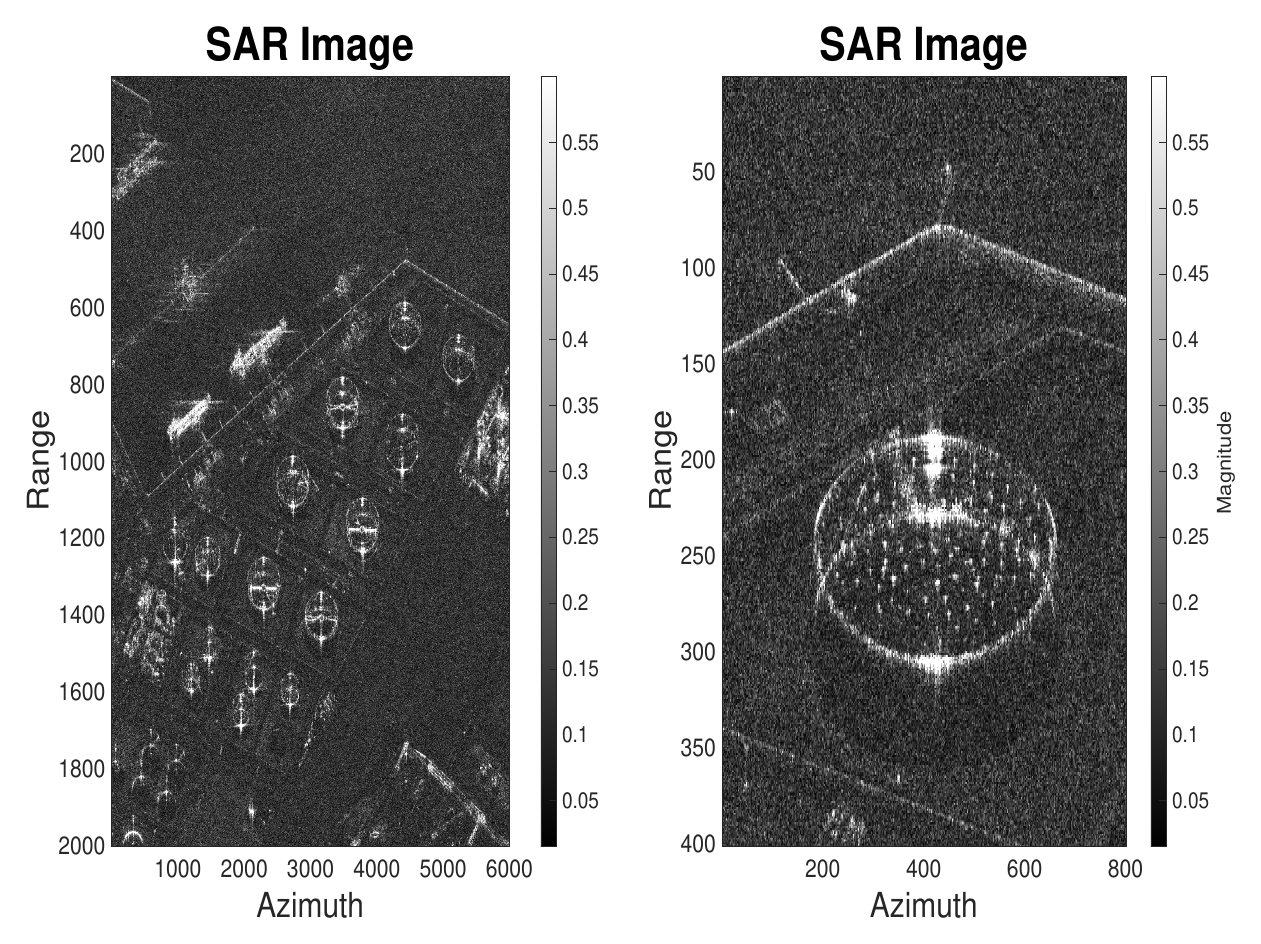}
    \caption{Zoomed SAR regions used for shadow, layover, and view-direction interpretation.}
    \label{fig:zoomed-sar}
\end{figure}

SAR images are strongly affected by side-looking geometry. Bright layover responses occur when the radar-facing side of an elevated or angular object is mapped toward the sensor, while radar shadows occur where no direct illumination reaches the ground behind the object. The zoomed regions in \Fig{fig:zoomed-sar} contain compact bright responses and adjacent dark areas associated with industrial objects. Since shadows appear on the side opposite to illumination, the visible shadow orientation indicates that the radar illumination arrives from the upper northern side under the adopted image display convention. This qualitative inference is useful for separating actual low-backscatter water or asphalt from radar shadow and for interpreting displaced bright responses near tanks, quay structures, and industrial edges.

\subsection{Resolution Formulation and Spectral Assessment}

\begin{figure}[!h]
    \centering
    \includegraphics[width=0.98\linewidth]{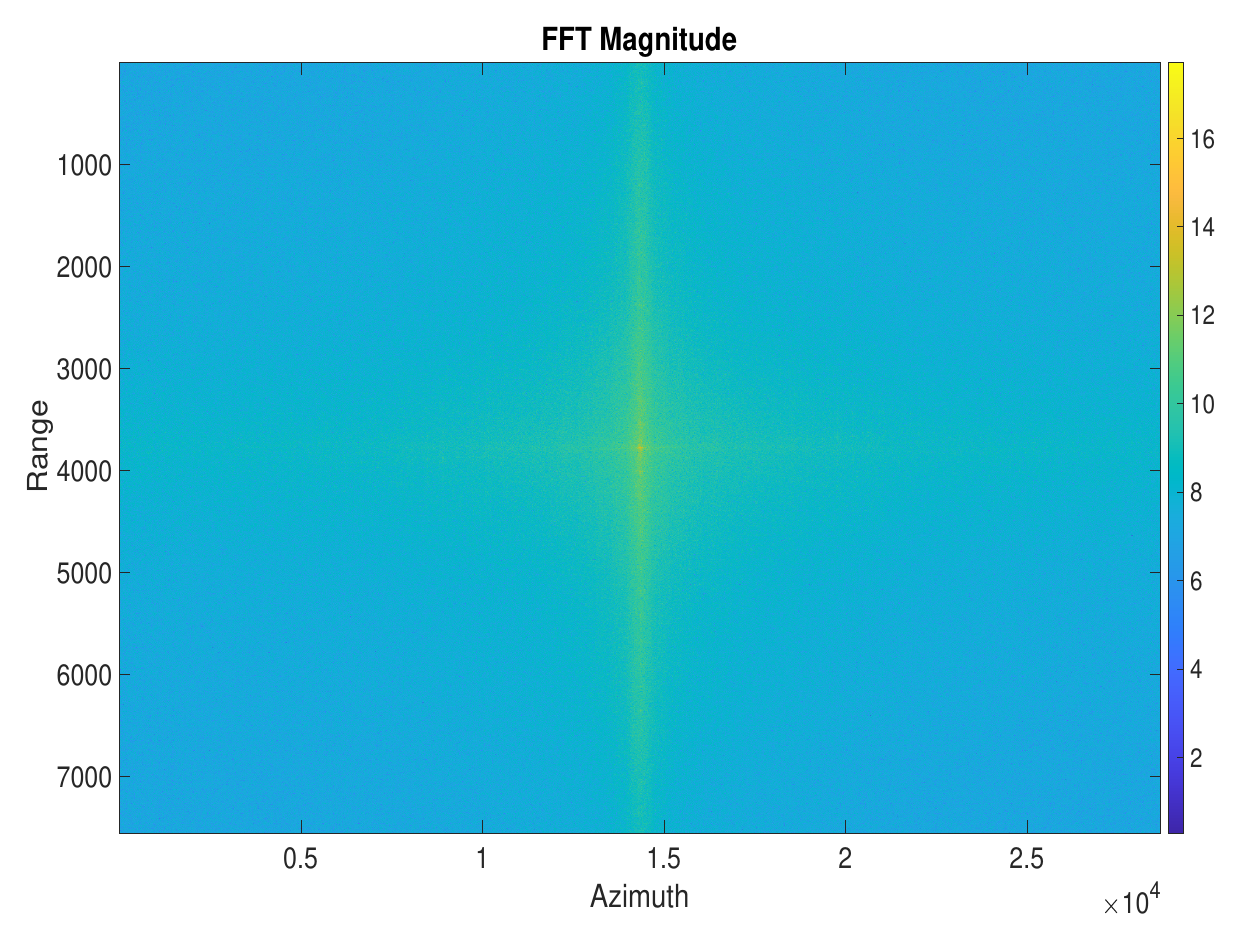}
    \caption{Magnitude of the two-dimensional FFT used for frequency-content-based resolution assessment.}
    \label{fig:fft}
\end{figure}

Range resolution follows from the ability to separate two echoes with different two-way propagation delays. If two targets differ by slant-range distance $\Delta R$, the delay difference is $\Delta\tau=2\Delta R/c$, where $c$ is the speed of light. A transmitted waveform with bandwidth $B$ has an approximate delay resolution of $1/B$ after matched filtering. Equating $\Delta\tau$ to $1/B$ gives the slant-range resolution
\begin{equation}
\rho_{sr}\approx\frac{c}{2B} .
\label{eq:slant-resolution}
\end{equation}
For ground-range interpretation, the slant-range cell is projected onto the ground by the incidence angle $\theta_i$, yielding
\begin{equation}
\rho_{gr}\approx\frac{c}{2B\sin\theta_i} .
\label{eq:ground-resolution}
\end{equation}
Azimuth resolution is governed by coherent aperture synthesis. In Doppler terms, if the focused data use an effective Doppler bandwidth $B_D$ and the platform velocity is $v$, the azimuth separability can be approximated by
\begin{equation}
\rho_{az}\approx\frac{v}{2B_D} .
\label{eq:az-doppler}
\end{equation}
For a focused stripmap SAR with physical antenna length $L_a$, the processed aperture commonly leads to the well-known approximation $\rho_{az}\approx L_a/2$ under ideal focusing. When complete sensor metadata are unavailable, the image spectrum can provide an empirical indicator of spatial-frequency support. The two-dimensional Fourier transform of the magnitude image is
\begin{equation}
\mathcal{A}(f_r,f_a)=\sum_{m=0}^{N_r-1}\sum_{n=0}^{N_a-1}A(m,n)e^{-j2\pi\left(f_r m+f_a n\right)} .
\label{eq:fft2}
\end{equation}
The width of the dominant spectral support in range and azimuth is used as a practical resolution indicator. The FFT magnitude in \Fig{fig:fft} supports approximate indicators of 0.42 m in range and $0.19^{\circ}$ in azimuth-angle separation for the available image-coordinate calibration. These values should be interpreted as empirical image-resolution estimates rather than as a replacement for metadata-based sensor characterization.

\subsection{Multitemporal Amplitude-Change Formulation}

For the geocoded pair, the master and slave amplitudes are written as $A_M=|M|$ and $A_S=|S|$. SAR intensity and amplitude observations include speckle, which may be described through a multiplicative model in which the measured amplitude is the product of a deterministic backscatter term and a stochastic granular component. A median operator over a spatial window $\Omega_p$ centered at pixel $p$ is used to reduce isolated fluctuations:
\begin{equation}
\widetilde{A}(p)=\operatorname{median}\{A(q):q\in\Omega_p\} .
\label{eq:median}
\end{equation}
Median filtering preserves edges better than simple averaging in many high-contrast areas, although it can attenuate isolated point scatterers. Therefore, the window size must be selected according to the required balance between smoothness and feature preservation.

\begin{figure}[!h]
    \centering
    \includegraphics[width=0.94\linewidth]{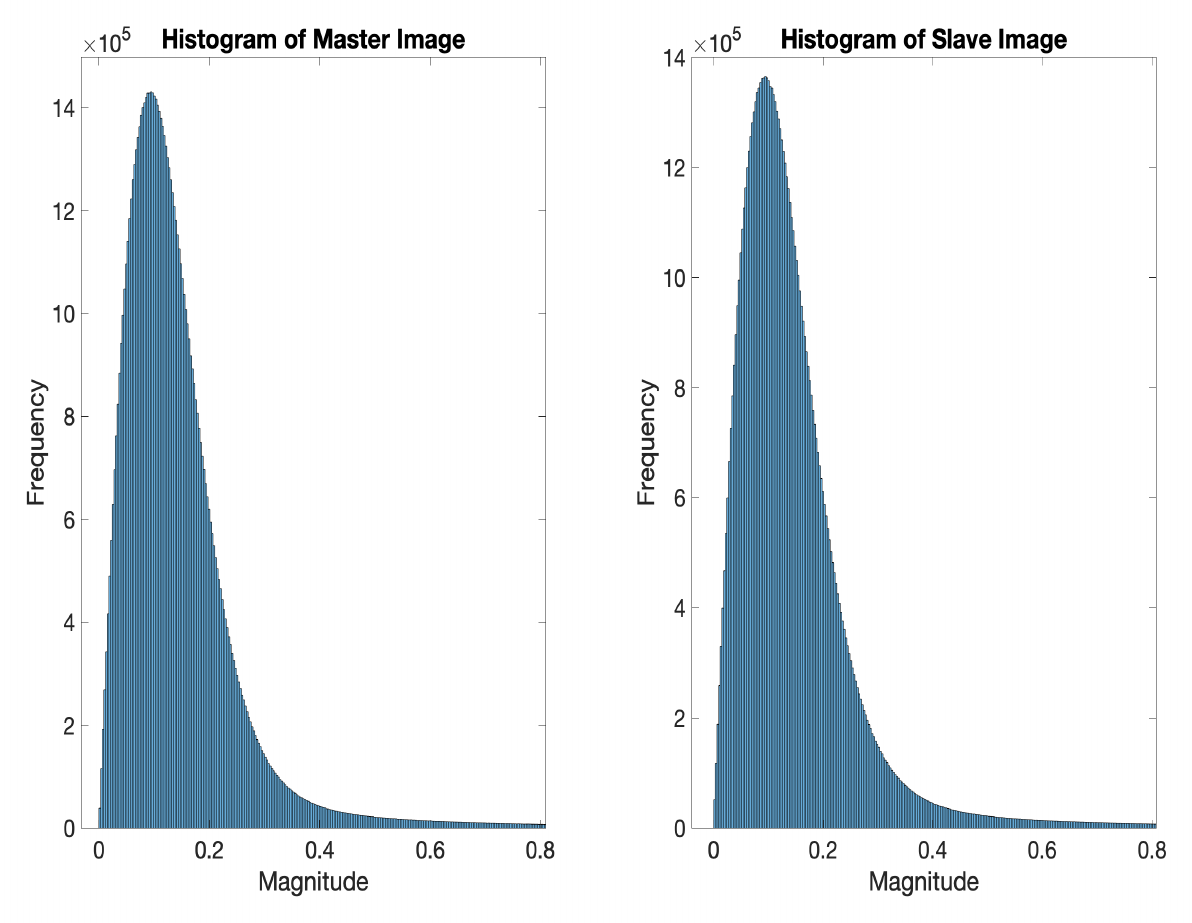}
    \caption{Amplitude histograms of the master and slave SAR images.}
    \label{fig:master-slave-histograms}
\end{figure}

\begin{figure}[!b]
    \centering
    \includegraphics[width=0.98\linewidth]{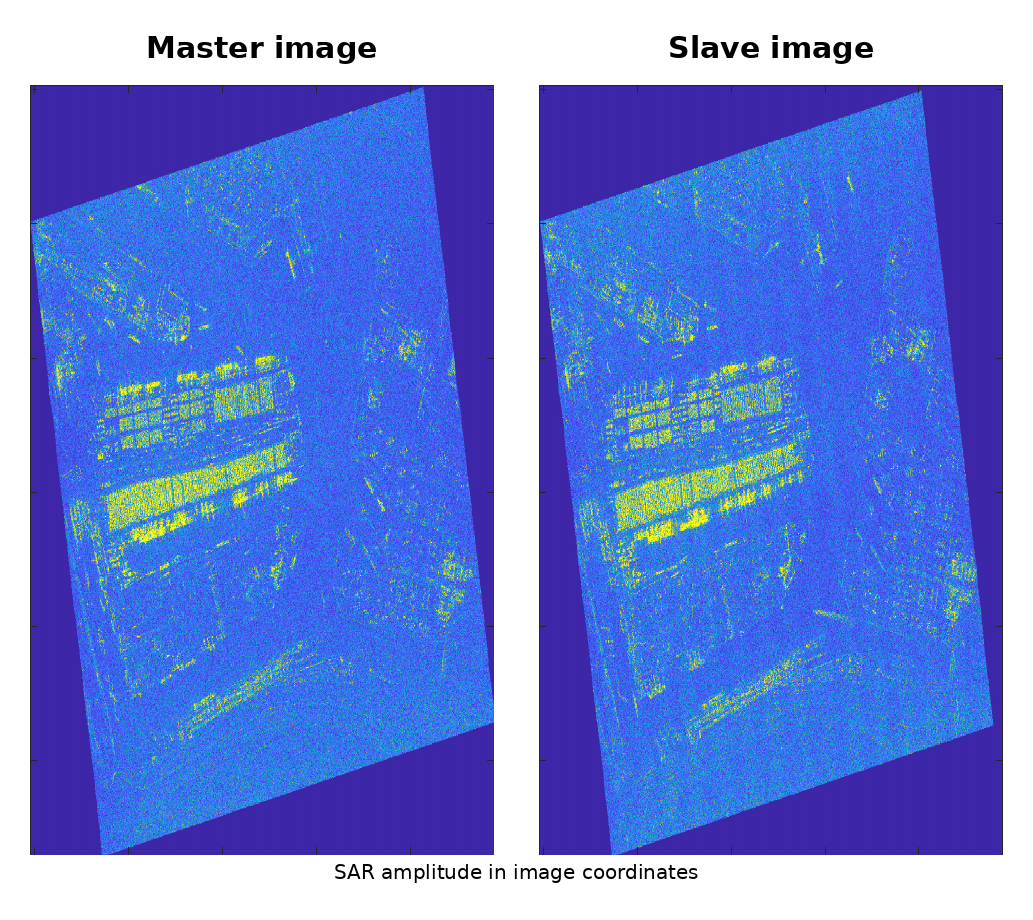}
    \caption{Geocoded master and slave SAR amplitude images displayed with common scaling.}
    \label{fig:master-slave}
\end{figure}

The primary amplitude-change layer is computed as the absolute difference between filtered master and slave amplitudes:
\begin{equation}
D(p)=\left|\widetilde{A}_M(p)-\widetilde{A}_S(p)\right| .
\label{eq:amplitude-difference}
\end{equation}
A normalized alternative can be useful when radiometric scaling differs between acquisitions:
\begin{equation}
D_N(p)=\frac{\left|\widetilde{A}_M(p)-\widetilde{A}_S(p)\right|}{\widetilde{A}_M(p)+\widetilde{A}_S(p)+\epsilon},
\label{eq:normalized-difference}
\end{equation}
where $\epsilon$ prevents division by zero. The present visualization focuses on \Eq{eq:amplitude-difference} because the master and slave products are displayed under a common interval. The histograms in \Fig{fig:master-slave-histograms} motivate the common scaling, while \Fig{fig:master-slave} shows the paired SAR amplitudes. The exported geospatial products are displayed in \Fig{fig:qgis-master-slave}, \Fig{fig:qgis-filtered}, and \Fig{fig:change-map}.

\subsection{Interferometric Coherence Formulation}

Interferometric coherence measures the normalized complex correlation between two co-registered SAR observations. The population form is
\begin{equation}
\gamma=\frac{\E\left[MS^{*}\right]}{\sqrt{\E\left[|M|^2\right]\E\left[|S|^2\right]}} ,
\label{eq:coherence}
\end{equation}
where $S^{*}$ is the complex conjugate of the slave image. The normalization removes the direct influence of signal power and constrains the coherence magnitude to the interval $0\leq |\gamma|\leq 1$ under ideal estimation. In finite images, the expectation is replaced by local spatial averaging within a window $\Omega_p$, producing
\begin{equation}
\hat{\gamma}(p)=\frac{\sum\limits_{q\in\Omega_p}M(q)S^{*}(q)}{\sqrt{\sum\limits_{q\in\Omega_p}|M(q)|^2\sum\limits_{q\in\Omega_p}|S(q)|^2}} .
\label{eq:sample-coherence}
\end{equation}
The numerator preserves the complex phase relationship, while the denominator normalizes by the local energy of both acquisitions. Low coherence can result from physical surface changes, moving vessels, water motion, vegetation or roughness variation, thermal noise, temporal decorrelation, or imperfect co-registration. Large windows reduce estimator variance but blur spatial transitions; small windows preserve detail but produce noisier estimates. Window sizes of $3\times3$, $5\times5$, $7\times7$, and $9\times9$ pixels are therefore examined.

\vspace{4pt}
\section{Results and Discussion}
\vspace{4pt}
\subsection{Amplitude Products and Imaging Geometry}
\vspace{4pt}
The SAR magnitude image in \Fig{fig:sar-image} confirms the high dynamic range expected from a dense port scene. Bright signatures are concentrated along linear infrastructure, storage-tank boundaries, quay edges, ship-like targets, and angular industrial elements. The histogram in \Fig{fig:histogram} shows that most pixels occupy low-to-moderate amplitude values, which explains why unrestricted display scaling would obscure relevant features. The grayscale image in \Fig{fig:gray-sar} provides a clearer representation of compact scatterers and elongated port structures after the selected display interval is applied.

\begin{figure}[!h]
    \centering
    \includegraphics[width=0.97\linewidth]{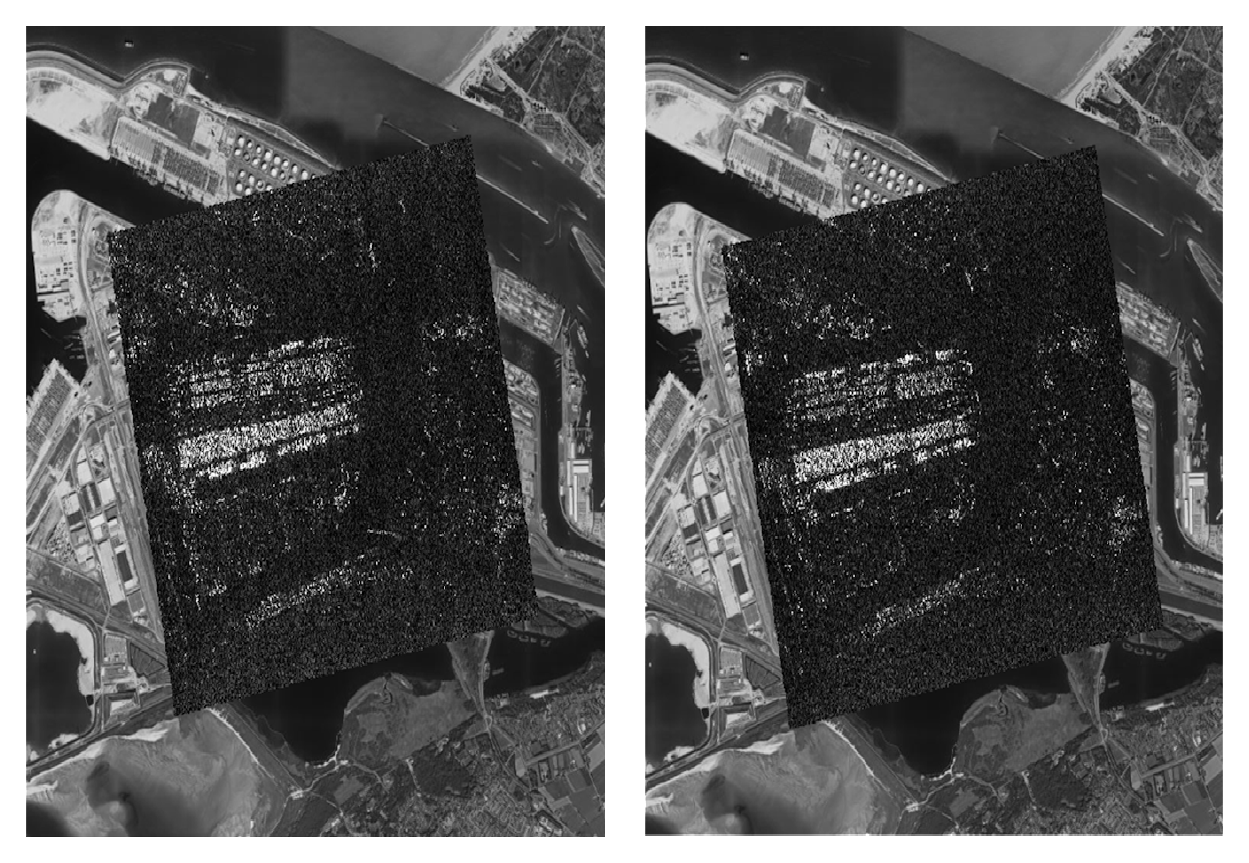}
    \caption{Master and slave SAR images inspected after GeoTIFF export.}
    \label{fig:qgis-master-slave}
\end{figure}

The zoomed examples in \Fig{fig:zoomed-sar} show that qualitative geometric interpretation remains valuable even without complete sensor metadata. A bright circular or cylindrical response accompanied by a neighboring shadow provides evidence of the illumination direction. In the port context, this cue helps distinguish shadowed land surfaces from low-backscatter water and supports the interpretation of strong responses near tanks or vertical metallic structures. The spectral result in \Fig{fig:fft} indicates dominant energy concentrated around the main spatial-frequency axes. The empirical resolution indicators are consistent with the visual presence of compact bright structures, but the estimates should be treated as scene- and processing-dependent because the FFT is applied to the displayed SAR magnitude rather than to a fully characterized sensor point-spread function.

\subsection{Amplitude-Based Change Detection}

The master and slave histograms in \Fig{fig:master-slave-histograms} have similar overall shapes, supporting the use of a shared visualization range. The paired images in \Fig{fig:master-slave} contain stable large-scale structures but also localized amplitude differences around waterfront infrastructure and high-backscatter industrial zones. These differences may correspond to operational activity, vessel displacement, water-surface variation, changes in storage or cargo areas, or viewing-geometry interactions with small targets.

\begin{figure}[!h]
    \centering
    \includegraphics[width=0.97\linewidth]{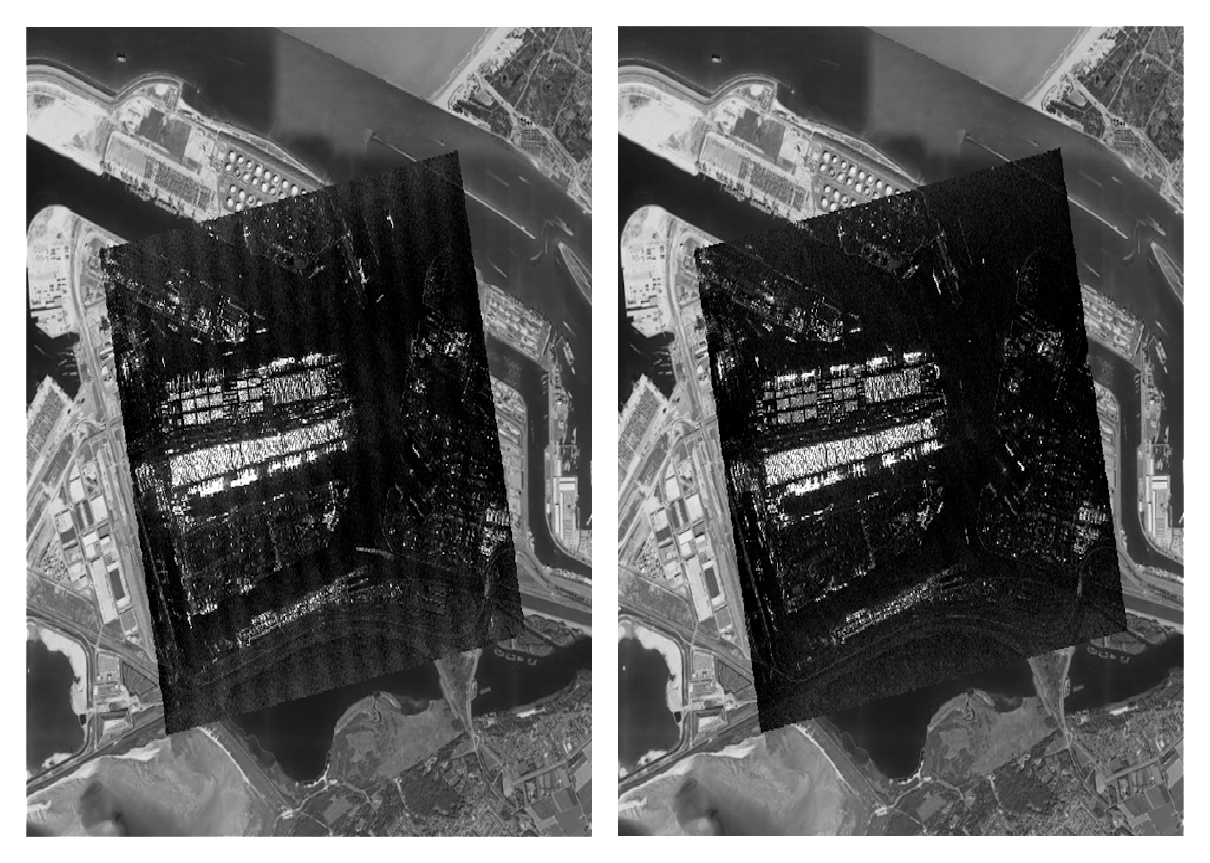}
    \caption{Median-filtered master and slave SAR images after geospatial export.}
    \label{fig:qgis-filtered}
\end{figure}

\begin{figure}[!h]
    \centering
    \includegraphics[width=0.98\linewidth,height=0.6\linewidth]{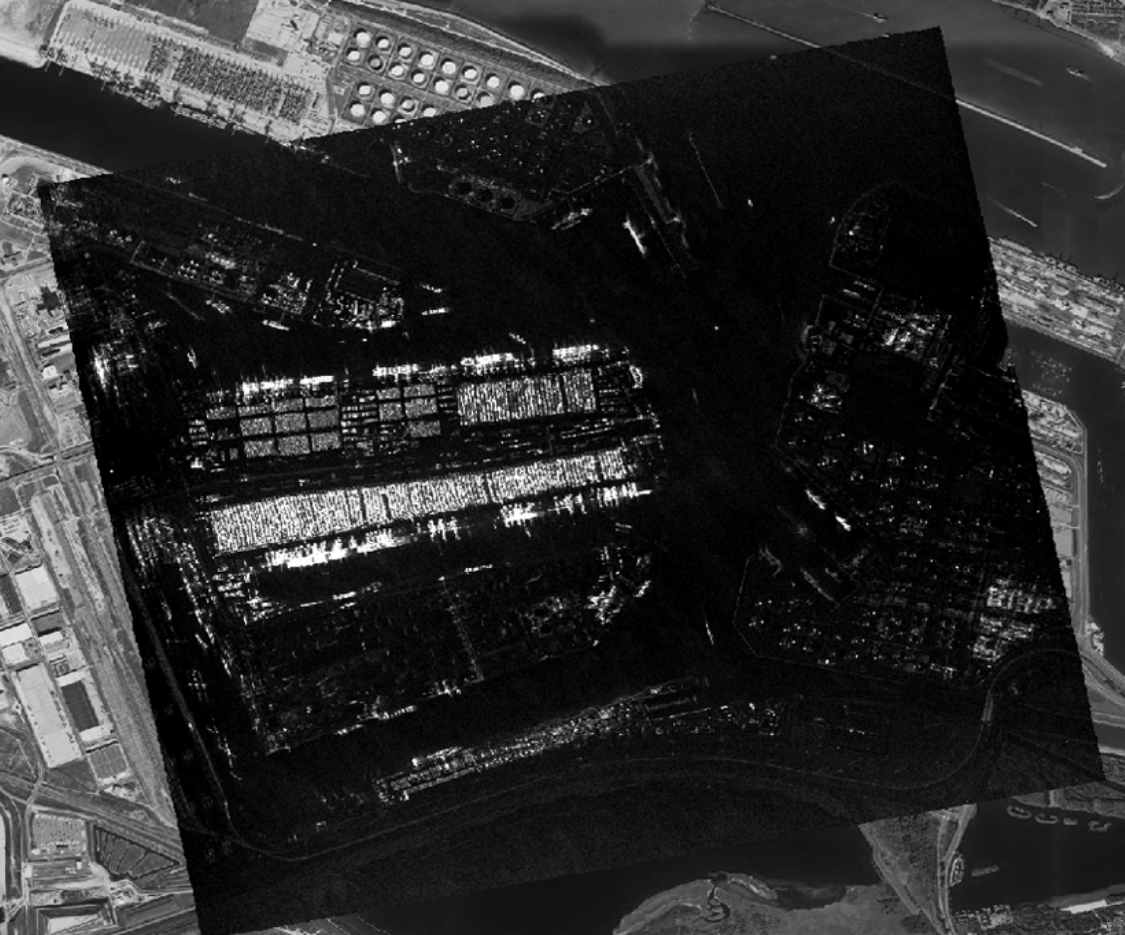}
    \caption{Filtered amplitude change map visualized in the geographic information system environment. Bright regions indicate stronger magnitude differences between the master and slave images.}
    \label{fig:change-map}
\end{figure}

The GeoTIFF inspection in \Fig{fig:qgis-master-slave} confirms that the processed SAR products retain their spatial referencing after export. The median-filtered products in \Fig{fig:qgis-filtered} reduce granular speckle and make broad patterns easier to compare. Nevertheless, the filtering operation also suppresses isolated bright points that may represent meaningful port objects. The resulting change map in \Fig{fig:change-map} should therefore be interpreted as a qualitative amplitude-change layer rather than as a fully validated classification map. Strong responses in this layer identify candidate change regions that would require threshold selection, radiometric calibration, and independent validation for operational deployment.

\begin{figure}[!h]
    \centering
    \includegraphics[width=0.98\linewidth]{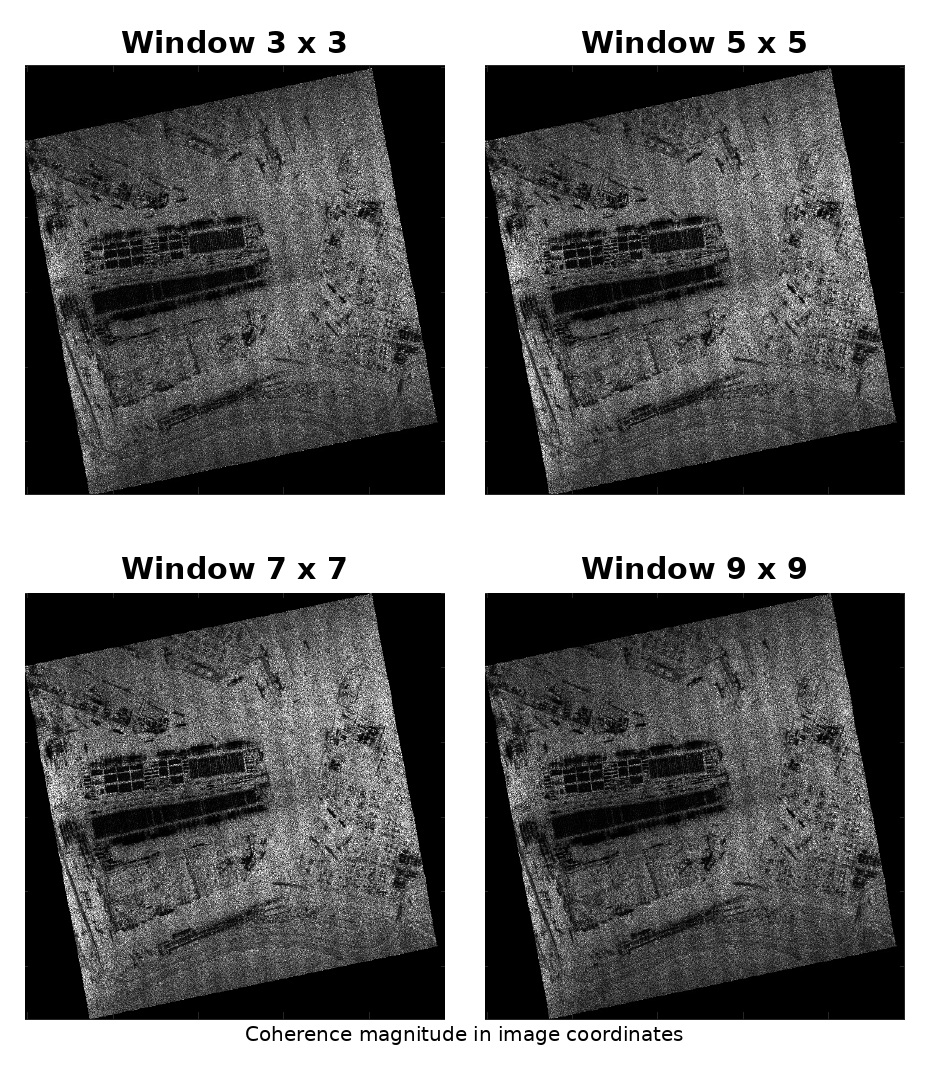}
    \caption{Interferometric coherence estimated with multiple spatial averaging window sizes.}
    \label{fig:coherence-windows}
\end{figure}

\subsection{Coherence and Spatial Averaging Behavior}

The coherence maps in \Fig{fig:coherence-windows} demonstrate the expected trade-off between spatial detail and estimator stability. The $3\times3$ window preserves fine structure but exhibits stronger local variability. Increasing the window size to $5\times5$, $7\times7$, and $9\times9$ progressively smooths the image, reduces random fluctuations, and widens transitions between coherent and decorrelated areas. The most appropriate window therefore depends on whether the analysis prioritizes sharp localization of small port features or robust identification of broader stable zones.

\begin{figure}[!h]
    \centering
    \includegraphics[width=0.98\linewidth]{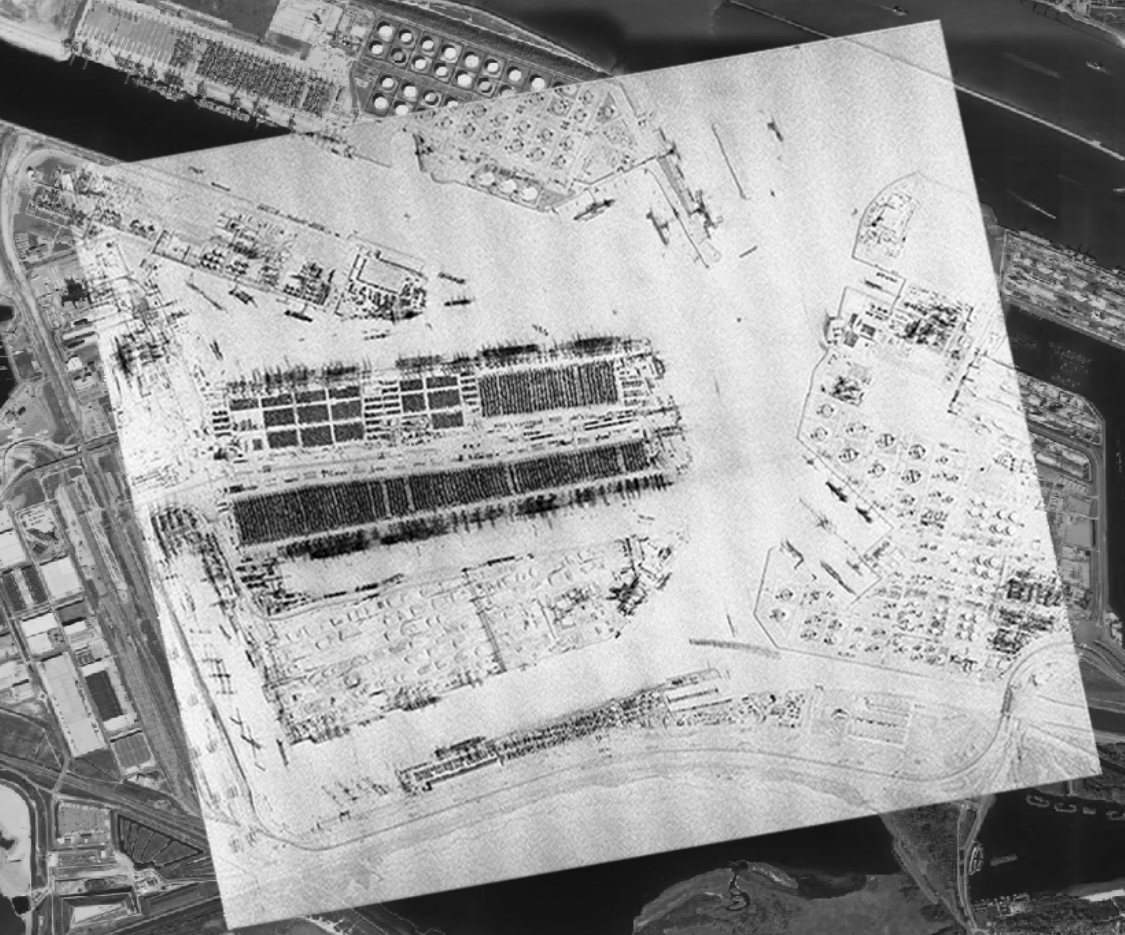}
    \caption{Final interferometric coherence product generated from the multitemporal SAR pair.}
    \label{fig:interferometric-coherence}
\end{figure}

The final coherence product in \Fig{fig:interferometric-coherence} complements the amplitude-change map by measuring temporal consistency of the complex scattering process. Stable infrastructure with persistent geometry tends to maintain higher coherence, whereas water surfaces, moving vessels, operational rearrangement, and decorrelated industrial areas tend to produce lower coherence. Joint interpretation of \Fig{fig:change-map} and \Fig{fig:interferometric-coherence} is preferable because an amplitude increase or decrease does not necessarily imply phase stability, and a low-coherence region may arise even when the mean amplitude remains similar across acquisitions.

\section{Conclusion}

A formal SAR processing workflow has been presented for maritime port monitoring using complex SAR imagery, geocoded multitemporal products, and an optical reference image. Histogram-guided amplitude scaling improves the visibility of high-dynamic-range port features. Zoomed inspection of shadows and bright scatterers supports qualitative interpretation of radar illumination geometry. The resolution formulation links range separability to waveform bandwidth and azimuth separability to Doppler support, while the image-spectrum analysis provides empirical indicators for the available data. Multitemporal amplitude differencing, median filtering, GeoTIFF export, and coherence estimation jointly support the identification of candidate change regions and stable infrastructure.

Amplitude and coherence products provide complementary information. Amplitude differences identify changes in backscatter strength, whereas coherence measures the temporal stability of the complex scattering response. Median filtering improves visual clarity but should be selected carefully because excessive smoothing can weaken isolated industrial scatterers and small vessel-related signatures. Further development would benefit from radiometric calibration, terrain or ellipsoid correction, adaptive thresholding, multi-looking strategies, polarimetric descriptors, and validation against independent port activity records or high-resolution optical observations.

\end{document}